\documentclass[aps,prb,floatfix,twocolumn,superscriptaddress,showpacs,amsmath,amssymb]{revtex4}
\usepackage{graphicx}
\usepackage{epsfig} 
\usepackage{color}
\usepackage{ulem}

\allowdisplaybreaks
\newcommand{\ca}{c^{\phantom{\dagger}}}

\newcommand{\be}{\begin{equation}}
\newcommand{\ee}{\end{equation}}
\newcommand{\bea}{\begin{eqnarray}}
\newcommand{\eea}{\end{eqnarray}}
\newcommand{\ba}{\begin{eqnarray*}}
\newcommand{\ea}{\end{eqnarray*}}
\newcommand{\dagga}{{\phantom{\dagger}}}
\newcommand{\bR}{\mathbf{R}}
\newcommand{\bQ}{\mathbf{Q}}

\newcommand{\bk}{\mathbf{k}}

\newcommand{\dis}{\displaystyle}

\newcommand{\up}{\uparrow}
\newcommand{\down}{\downarrow}
\newcommand{\fract}[2]{\frac{\dis #1}{\dis #2}}
\newcommand{\Tr}{\mathrm{Tr}}
\newcommand{\eqn}[1]{(\ref{#1})}
\newcommand{\m}[1]{\mathcal{#1}} 
\newcommand{\eps}{\varepsilon}

\newcommand{\ket}[1]{| #1 \rangle}

\begin{document}
\title{Non-equilibrium dynamics in the antiferromagnetic Hubbard model} 

\author{Matteo Sandri}
\affiliation{
International School for Advanced Studies (SISSA), and CRS Democritos, CNR-INFM, - Via Bonomea 265, I-34136 Trieste, Italy} 
\author{Michele Fabrizio}
\affiliation{
International School for Advanced Studies (SISSA), and CRS Democritos, CNR-INFM, - Via Bonomea 265, I-34136 Trieste, Italy} 

\date{\today} 


\begin{abstract}
We investigate by means of the time-dependent Gutzwiller variational approach the out-of-equilibrium dynamics of an antiferromagnetic state evolved with the Hubbard model Hamiltonian after a sudden change of the repulsion strength $U$.  We find that magnetic order survives more than what expected on the basis of thermalization arguments, in agreement with recent DMFT calculations. In addition, we find evidence of a dynamical transition for quenches to large values of $U$ between a coherent antiferromagnet characterized by a finite quasiparticle residue to an incoherent one with vanishing residue, that finally turns into a paramagnet for even larger $U$. 
\end{abstract}
\maketitle

\section{Introduction}
In the last years the out of equilibrium physics in correlated systems has attracted considerable interest,  mainly driven by impressive experimental progresses. On one hand, trapped cold atoms, known to effectively realize simple model Hamiltonians, have been successfully exploited to investigate quench dynamics or field-driven non-equilibrium phenomena in quasi-isolated quantum many-body systems\cite{Bloch}. 
On the other hand, time-resolved femtosecond spectroscopies have made it possible to perturb solid state systems and access the dynamics of the electronic degrees of freedom before they thermalize with the environment and even before they equilibrate with the lattice\cite{Cavalieri}.
Overall, these experiments allow to study how strong correlation affects the out-of-equilibrium physics and possibly identify ``novel phases'' that cannot be reached by conventional thermal pathways.\\
To this end, a fundamental issue to address is the real time dynamics across a phase transition in which symmetry is broken or restored. The ultra-fast melting and creation of long range order in transition metal compounds has already been investigated in many experiments.\cite{Perfetti,Beaud,Fausti} On the theoretical side, however, while an equilibrium phase transition is a well established concept, there is yet no clear extension to the out of equilibrium case.\cite{Giacomo} The common viewpoint is that the initial excess energy $\Delta E$ 
turns into heat, hence the system evolves into a thermal state at a higher effective temperature $T_*$, higher the bigger $\Delta E$. Should $T_*$ exceed the critical temperature for a order-to-disorder phase transition, 
the system would dynamically disorder though initially ordered.\\

Recently the dynamics of a symmetry breaking state has been addressed by means of time-dependent DMFT in the single-band repulsive Hubbard model on a Bethe lattice.\cite{Werner_1,Werner_2} Such model, which may be considered as the simplest idealization of strongly correlated electrons, displays at equilibrium 
a N\'eel transition from a low temperature antiferromagnet (AFM) to a high temperature 
paramagnet (PM).  As mentioned, upon sudden changing the interaction strength, $U_i\rightarrow U_f$, one could dynamically move around the phase diagram and eventually cross the N\'eel transition. Refs. \onlinecite{Werner_1} and \onlinecite{Werner_2} showed that both for $U_f < U_i$ and $U_f>U_i$, long-lived non-thermal ordered states exist even though their expected $T_*$ is above the N\'eel temperature $T_N$. 
Moreover, it was found that for $U_f < U_i$, the melting of the AFM order is related to the existence of a non-thermal critical point with an associated vanishing amplitude mode. Both these features are consequence of pure non-equilibrium effects.

Here we address the same model dynamics by means of the time dependent Gutzwiller variational approach introduced in Ref. \onlinecite{Schiro&Fabrizio_prl}. This method, although being less accurate than DMFT, is computationally far less expensive and has already proved its reliability in reproducing the main results of DMFT in the out-of-equilibrium dynamics of paramagnetic states.
\cite{Schiro&Fabrizio_prl,Schiro&Fabrizio_prb,Sandri_lin} We find that also in the broken-symmetry dynamics, the time-dependent Gutzwiller tecnique correctly reproduces both the presence of a critical point at which magnetism disappears as well as the existence of non-thermal ordered states. Moreover, we find evidence of an additional critical point at $U_f>U_i$ between two antiferromagnetic states that we interpret as the magnetic analogue of a dynamical Mott transition. 

The paper is organized as follows. In section \ref{Time dependent Gutzwiller} we briefly present how the method works in the specific case of an antiferromagnet. In section \ref{Interaction quench} we move to discuss the results of a quench from an initial magnetic state, ground state of the Hamiltonian at repulsion $U_i$, evolved with the Hamiltonian at a different value $U_f$, both for $U_f<U_i$, section \ref{$U_f < U_i$ quench}, and $U_f>U_i$, section \ref{$U_f > U_i$ quench}. Finally, section \ref{Concluding remarks} 
is devoted to conclusions. 

\section{Time dependent Gutzwiller}
\label{Time dependent Gutzwiller}
In this section we briefly show how the time-dependent Gutzwiller technique introduced in Ref.  \onlinecite{ Schiro&Fabrizio_prl} has to be modified to treat the AFM dynamics within the single band Hubbard model at half filling, with Hamiltonian 
\be \label{HUBHAM}
\m{H} = - \sum_{\langle \bR,\bR' \rangle,\sigma } \Big(c^\dagger_{\bR\sigma} \ca_{\bR'\sigma} + H.c.\Big)  + \frac{U(t)}{2} \sum_{\bR} (n_{\bR} -1)^2,
\ee
where $c_{\bR\sigma}$ annihilates a spin-$\sigma$ electron at site $\bR$, $U(t)$ is the (time dependent) interaction strength and $n_{\bR} = \sum_\sigma c^\dagger_{\bR \sigma} \ca_{\bR \sigma} $. The hopping parameter is set equal to one and is our unit of energy.\\
We follow the same notations as Ref. \onlinecite{Hvar}, which the reader is referred to for a more detailed derivation.\\ 
The main idea of the time dependent Gutzwiller technique is to approximate the evolving wavefunction $\ket{\Psi(t)}$ in terms of a variational wavefunction whose dynamics is set by requiring the stationarity of the real time action
\be \label{LAG}
\m{L}(t) = \int_0^t d\tau \, \langle \Psi(\tau) | i \partial_\tau - \m{H}(\tau) \ket{\Psi(\tau)}.
\ee
In the same spirit of the ground state Gutzwiller method, one introduces the following ansatz for the evolving wavefunction\cite{Schiro&Fabrizio_prl}
\be \label{GWF}
\ket{\Psi(t)} = \prod_{\bR} \m{P}_{\bR}(t) \;\ket {\psi(t)}
\ee 
where $\ket {\psi(t)} $ is a generic time-dependent variational Slater determinant, and $\m{P}_{\bR}(t)$ a time-dependent variational local operator.\\
Upon introducing a basis for the local Fock space
\be \label{BASE}
\ket{ \bR, \{n\} } = \prod_{\alpha=\uparrow,\downarrow} (c^\dagger_{\bR \alpha})^{n_\alpha}
\ee
one can parametrize the Gutzwiller projector in terms of a set of time dependent variational parameters $\Phi_{\bR \;\{n\}}(t)$
\be
\m{P}_{\bR}(t) = \sum_{\{ n\}} \frac{\Phi_{\bR \; \{n\}}(t)}{\sqrt{P^{(0)}_{\bR \; \{n\} }(t) }} \ket {\bR, \{n\}}\langle {\bR, \{n\}} |
\ee
where
\be
P^{(0)}_{\bR \; \{n\} }(t) = \langle \psi(t) \ket {\bR, \{n\}} \langle {\bR, \{n\}} \ket{\psi(t)}.
\ee
In Ref. \onlinecite{Hvar} it was shown that the stationarity of (\ref {LAG}) amounts to solve a set of coupled differential equations that determine the evolution of the uncorrelated wavefunction $\ket{\psi(t)}$ and the variational parameters $\Phi_{\bR \;\{n\}}(t)$:
\bea 
& &i \partial_t \ket{\psi(t)} = \m{H}_*[\hat{\Phi}(t)] \; \ket{\psi(t)} \label{TE1}\\
& &i \partial_t \hat{\Phi}_{\bR}(t)  = \hat{U}(t) \hat{\Phi}_{\bR}(t) + \langle \psi(t) | \frac{\partial \m{H}_* [\hat{\Phi}(t)]}{\partial \hat{\Phi}^\dagger_{\bR}(t) } \ket{\psi(t)}. \label{TE2}
\eea
With the notation $\hat{O}_\bR$ we indicate the matrix representation of the operator $O_\bR$ on the Fock basis (\ref{BASE}). If we assume the magnetization directed along $z$, then we can choose  $\hat{\Phi}_{\bR}$ to be a diagonal matrix with diagonal elements $\Phi_{\bR \;\{0\}}$, for empty site, $\Phi_{\bR \;\{\uparrow \}}$ and $\Phi_{\bR \;\{\downarrow \}}$, for singly occupied site with a spin up or down electron, respectively, and finally $\Phi_{\bR \;\{\uparrow \downarrow \}}$ for a doubly occupied site.\\
The Slater determinant evolves according to a ``renormalized'' one-body Hamiltonian 
\be \label{RH}
\m{H}_*[\hat{\Phi}(t)] = -\sum_{\langle \bR,\bR' \rangle,\sigma } 
\Big(R^*_{\bR \sigma}(t) c^\dagger_{\bR\sigma} R_{\bR' \sigma}(t) \ca_{\bR'\sigma} + H.c.\Big) 
\ee
which is self-consistently coupled to the evolution of the matrix $\hat{\Phi}_{\bR}(t)$ through the renormalization factors
\be \label{RENO}
R_{\bR \sigma}(t) = \frac{1}{\sqrt{n_{\bR \sigma}(t) (1-n_{\bR \sigma}(t)) }}
\Tr (\hat{\Phi}_{\bR}^\dagger(t) \hat{c}_{\bR \sigma} \hat{\Phi}_{\bR \sigma}(t) \hat{c}^\dagger_{\bR \sigma} ).
\ee

In the presence of N\'eel AFM order we can separate the bipartite lattice into two sublattices A and B such that Eq. (\ref{RH}) becomes 
\be \label{HAM_EFF}
\m{H}_*(t) = -\sum_{ \langle \bR_a, \bR_{\bar{a}}  \rangle,\sigma} \Big(R^*_{\bR_a \sigma}(t) R_{\bR_{a} -\sigma}(t) 
c^\dagger_{\bR_a\sigma} \ca_{\bR_{\bar{a}}\sigma} + H.c.\Big)
\ee
where if $a=A$ then $\bar{a}=B$ and vice versa, and we make use of 
\be \label{PL}
R_{\bR_a \sigma} = R_{\bR_{\bar{a}} -\sigma} \mbox{, with } a \in \{A,B\}.
\ee
It is more convenient to work in Fourier space where Eq. (\ref{HAM_EFF}) reads
\bea \label{HF}
\m{H}_*(t) = \sum_{\bk\sigma} \eps(\bk) \Big [ \Re{ \big( R^*_{\bR_A \sigma}(t) R_{\bR_A -\sigma}(t) \big)  } c^\dagger_{\bk \sigma} \ca_{\bk \sigma}  \nonumber \\
-i \Im{\big( R^*_{\bR_A \sigma}(t) R_{\bR_A -\sigma}(t) \big) } c^\dagger_{\bk \sigma} \ca_{\bk + \bQ \sigma} \Big ]
\eea
with $\eps(\bk) = \frac{1}{N} \sum_{ \langle \bR_a, \bR_{\bar{a}}  \rangle} e^{i\bk \cdot (\bR_a- \bR_{\bar{a}})}$ where $N$ is the number of sites, and the vector $\bQ$ such that 
\be
e^{i\bQ \cdot \bR_a} = 
\left \{	
\begin{array}{cc}
 1 & \mbox{if } a \in A \\
 -1 & \mbox{if } a \in B
\end{array}
\right. .
\ee

The time evolution of the uncorrelated $\ket{\psi(t)}$ can then be re-casted into that of $\Delta_{\bk \bk'}^\sigma(t) := \langle \psi(t)| c^\dagger_{\bk \sigma} \ca_{\bk' \sigma} \ket{\psi(t)}$ whose equations of motion are
\bea \label{SD1}
& i\partial_t \Delta_{\bk \bk}^\sigma  &= -i \eps(\bk) \Im{\big( Z^\sigma (t) \big)} 
\Big (  \Delta_{\bk \bk+\bQ}^\sigma + \Delta_{\bk+\bQ \bk}^\sigma  \Big ) \nonumber \\
& i\partial_t \Delta_{\bk \bk+\bQ}^\sigma &= -2\eps(\bk) \Re{\big( Z^\sigma (t) \big)} \Delta_{\bk \bk+\bQ}^\sigma  \\
& & + i \eps(\bk) \Im{\big( Z^\sigma (t) \big)} \Big (  \Delta_{\bk \bk}^\sigma - \Delta_{\bk+\bQ \bk+\bQ}^\sigma  \Big ). \nonumber
\eea
To simplify notations we introduced the quantity $Z^\sigma (t) = R^*_{\bR_A \sigma}(t) R_{\bR_A -\sigma}(t)$. By construction it follows that 
\be \label{DENS}
n_{A(B) \sigma}(t) = \frac{1}{N}\sum_\bk \Delta_{\bk \bk}^\sigma(t) \pm \Delta_{\bk \bk+\bQ}^\sigma(t).
\ee 
The evolution of the uncorrelated wavefunction is self-consistently coupled to equation (\ref{TE2}) that, because of (\ref{PL}), can be evaluated for a single sublattice and reads
\bea \label{GP1}
i\frac{\partial \hat{\Phi}_A}{\partial t} &=& \hat{U} \hat{\Phi}_A(t)   \\ 
&+& \frac{1}{N}\sum_{k,\sigma} \eps(\bk) \Big[ R_{A -\sigma} \big ( \Delta_{\bk \bk}^\sigma(t) - \Delta_{\bk \bk+\bQ}^\sigma(t)  \big) \frac{\partial R^*_{A \sigma}}{\partial \hat{\Phi}^\dagger_A} \nonumber \\
&+& R^*_{A -\sigma} \big ( \Delta_{\bk \bk}^\sigma(t) + \Delta_{\bk \bk+\bQ}^\sigma(t)  \big) \frac{\partial R_{A \sigma}}{\partial \hat{\Phi}^\dagger_A} \Big ] . \nonumber
\eea
In conclusion Eqs. (\ref{SD1})-(\ref{GP1}) together with Eqs. (\ref{RENO}) and (\ref{DENS}) define a set of coupled non-linear differential equations which must be solved numerically.\\
In spite of the nonlinearity, the dynamics is still oversimplified and we do not expect to reach thermalization in the long time limit, mainly because the evolution of the Slater determinant still admits an infinite number of integrals of motion. In fact, the dynamics of $|\psi(t)\rangle$ does not mix different $(\bk,\bk+\bQ)$ subspaces. Within each subspace, the set of equations (\ref{SD1}) can be mapped onto the dynamics of a pseudospin-$\frac{1}{2}$ Hamiltonian. Indeed, upon defining 
\bea 
\Delta_{\bk \bk}^\sigma - \Delta_{\bk+\bQ \bk+\bQ}^\sigma & \equiv & \langle \sigma_1 \rangle \nonumber \\
\Delta_{\bk \bk+\bQ}^\sigma + \Delta_{\bk+\bQ \bk}^\sigma & \equiv & \langle \sigma_2 \rangle \nonumber \\
\Delta_{\bk \bk+\bQ}^\sigma - \Delta_{\bk+\bQ \bk}^\sigma & \equiv & {-i}\langle \sigma_3 \rangle \nonumber 
\eea 
(where in this case $\bk$ is restricted to the Magnetic Brillouin Zone (MBZ)), the set of equations (\ref{SD1}) is equivalent to solving the dynamics of the pseudo-spin Hamiltonian
\be \label{Hspin}
\m{H}^S_{\bk \sigma}(t) = \eps(\bk) \Im{\big( Z_\sigma (t) \big)} \sigma_3 - \eps(\bk) \Re{\big( Z_\sigma (t) \big)} \sigma_1
\ee
where $\sigma_{1,2,3}$ are Pauli matrices. 
Indeed, as we mentioned, the length of the pseudo-spin is a conserved quantity in each subspace.\\

It is generally believed that the average values of local operators along the unitary evolution 
of a wave function $|\Psi\rangle$, generically consisting of a superposition of a macroscopic number of eigenstates, will approach at long times the thermal averages on a Boltzmann-Gibbs distribution 
at an effective temperature $T_*$ for which the internal energy coincides with the energy of the wave function $|\Psi\rangle$, conserved during the unitary evolution, i.e.
\[
\fract{\text{Tr}\bigg(\text{e}^{-\mathcal{H}/T_*}\;\mathcal{H}\bigg)}
{\text{Tr}\bigg(\text{e}^{-\mathcal{H}/T_*}\bigg)} = \langle \Psi|\,\mathcal{H}\,|\Psi\rangle.
\]
Therefore it is worth comparing the results of the time-dependent Gutzwiller technique with equilibrium results at finite temperature obtained by a similar technique. For that purpose, we shall make use of an extension to finite temperature of the Gutzwiller variational approach recently proposed.\cite{Sandri_finT} In brief, the thermal values are computed minimizing the following variational estimate of the free energy,
\bea
F \leq && \, \min_{\{\rho_*,\hat{\Phi}\}}\Bigg\{  \sum_{\langle \bR,\bR' \rangle,\sigma }\Tr\bigg[
\rho_*\,
\Big(-R_{\bR \sigma} R_{\bR ' \sigma} c^\dagger_{\bR \sigma} c_{\bR ' \sigma}   \nonumber\\
&& +H.c.\Big) \bigg]
+ \sum_\bR\, \Tr\Big(\hat{\Phi}_\bR^\dagger\,\hat{U}\,\hat{\Phi}_\bR^\dagga\Big) \nonumber\\
&& \qquad \quad - T\,\text{Max}\Big(S_\text{var}\big(\rho_*,\hat{\Phi}^\dagger\hat{\Phi}\big) ,0\Big)
\Bigg\},\label{F-GW}
\eea
where $\rho_* = \text{e}^{-\beta\mathcal{H}_*}/(\Tr \, \text{e}^{-\beta\mathcal{H}_*})$ is the Boltzmann distribution corresponding to the variational Hamiltonian $\mathcal{H}_*$, and the variational estimate of the entropy reads
\bea
S_\text{var}\big(\rho_*,\hat{\Phi}^\dagger\hat{\Phi}\big) &=& -\Tr \Big(\rho_*\,\log\rho_*\Big)
\nonumber\\
&&  - \sum_{\bR,\{n\}} |\Phi_{\bR n}|^2\log\Big(\frac{|\Phi_{\bR n}|^2}{P^{(0)}_{\bR n}}\Big).  
\eea 
\\

We conclude this section remarking that all the above treatment is strictly variational 
only in the limit of infinite coordination number, where the exact averages on the Gutzwiller variational wavefunction (or the thermal averages on the variational canonical distribution) coincide with 
those we have computed.\cite{Hvar} However the approach remains essentially a mean-field one, hence, although improves the time-dependent Hartree-Fock approximation simply because of the larger number of variational parameters, it misses dissipative processes that in reality bring the system to a stationary state. 
In spite of that, the Gutzwiller approach seems to reproduce quite satisfactorily the main results obtained by exact DMFT calculations, whenever a comparison is possible and even when 
time-dependent Hartree-Fock fails completely, like in the case of quantum quenches within the paramagnetic 
sector.\cite{ Schiro&Fabrizio_prl}

In finite coordination lattices the approach is not anymore variational. Nevertheless, it is common to keep using the same expressions also in these more physical cases, which goes under the name of \textit{Gutzwiller Approximation}. Even though to our knowledge there are so far no exact out-of-equilibrium results to compare with in finite coordination lattices, recent high order perturbative calculations in one and two dimensions\cite{Hamerla1,Hamerla2} bring results quite similar to those obtained in Ref. \onlinecite{Schiro&Fabrizio_prl} through the Gutzwiller approach. \\
At equilibrium, instead, the Gutzwiller 
approximation seems to reproduce well exact variational Monte Carlo calculations on the Gutzwiller wave functions,\cite{Lanata-Kondo} and, when applied in combination with {\sl ab-initio} density functional theory methods, also physical properties of real materials.\cite{Lanata-Cerio}

\section{Interaction quench}
\label{Interaction quench}

\begin{figure}[!h]
\includegraphics[scale=0.32]{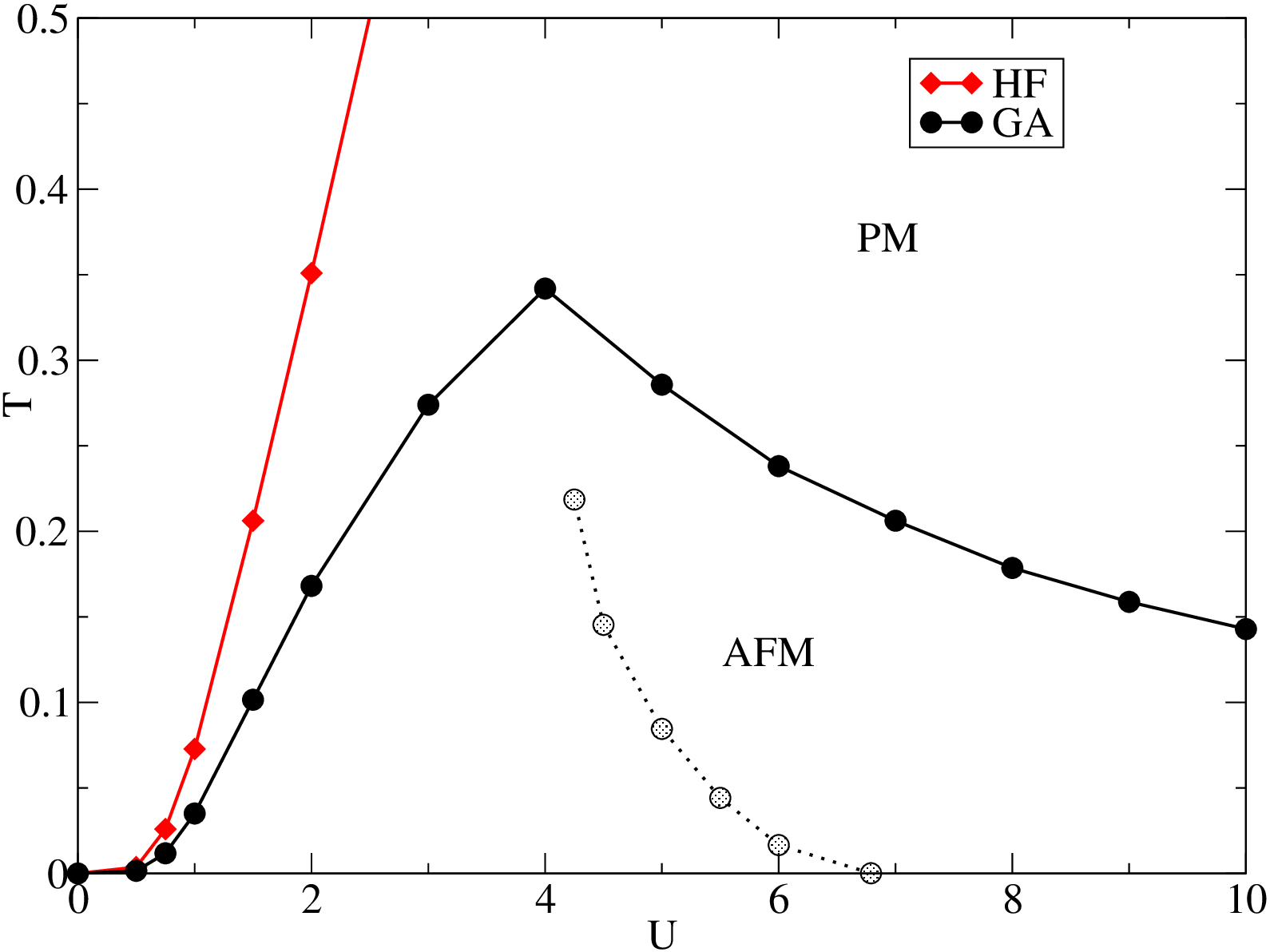}
\caption{(color online) Finite temperature phase diagram for the single band Hubbard model as obtained by mean of the finite temperature Gutzwiller approach. The solid black line separates the AFM solution from the PM phase. The dotted line indicates the MIT transition 
when only paramagnetic states are considered. 
The red line is the N\'eel temperature 
within the Hartree-Fock approximation.}
\label{fig:phaseFT}
\end{figure}

In this section we apply the time dependent Gutzwiller approach to study the dynamics of (\ref{HUBHAM}) after a sudden quench of the interaction strength, $U(t) = U_i + (U_f-U_i) \theta(t)$, where $\theta(t)$ is the Heaviside function.
Although an instantaneous quench is distant from the real practice in experiments, it is a well-controlled theoretical excitation protocol and suffices well the scope of this work. 
We assume nearest neighbor hopping on an infinitely branched  Bethe lattice, i.e. a semicircular density of states $D(\eps) = \sqrt{4-\eps^2}/(2\pi)$, in which case the Gutzwiller approximation becomes exact. We remark that the momentum representation we previously adopted is not appropriate for a Bethe lattice but can be easily extended in this case. \\
In Fig. \ref{fig:phaseFT} we plot the finite temperature phase diagram for the model as found by means of the finite temperature extension of the Gutzwiller technique.\cite{Sandri_finT} We see that the low temperature AFM ordered phase compares qualitatively well with the DMFT results.\cite{Werner_2}
In particular, the Gutzwiller wavefunction is able, unlike straight Hartree-Fock, to describe a finite temperature Mott insulating phase devoid 
of magnetism.

\subsection{$U_f < U_i$ quench}
\label{$U_f < U_i$ quench}

\begin{figure}[!h]
\includegraphics[scale=0.32]{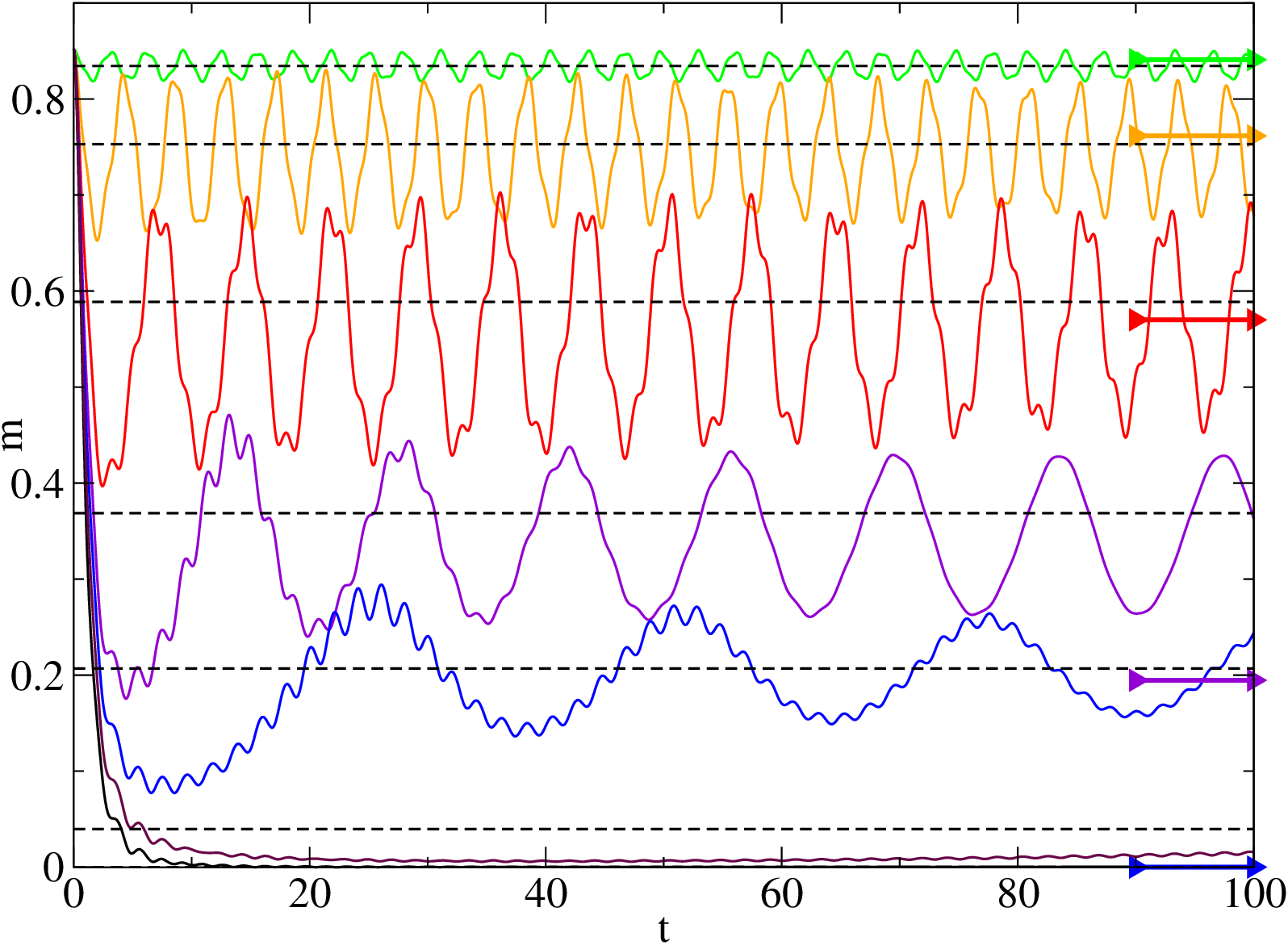}
\caption{(color online) Time evolution of the staggered magnetization $m$ for quenches $U_i=4.0 \rightarrow U_f = 3.8,3.2,2.6,2.2,2.0,1.8,1.6$. The bold arrows indicate the corresponding thermal values, 
$m_{th}$, while the black dashed lines indicate the long time averages.}
\label{fig:M}
\end{figure}

\begin{figure}[!h]
\includegraphics[scale=0.32]{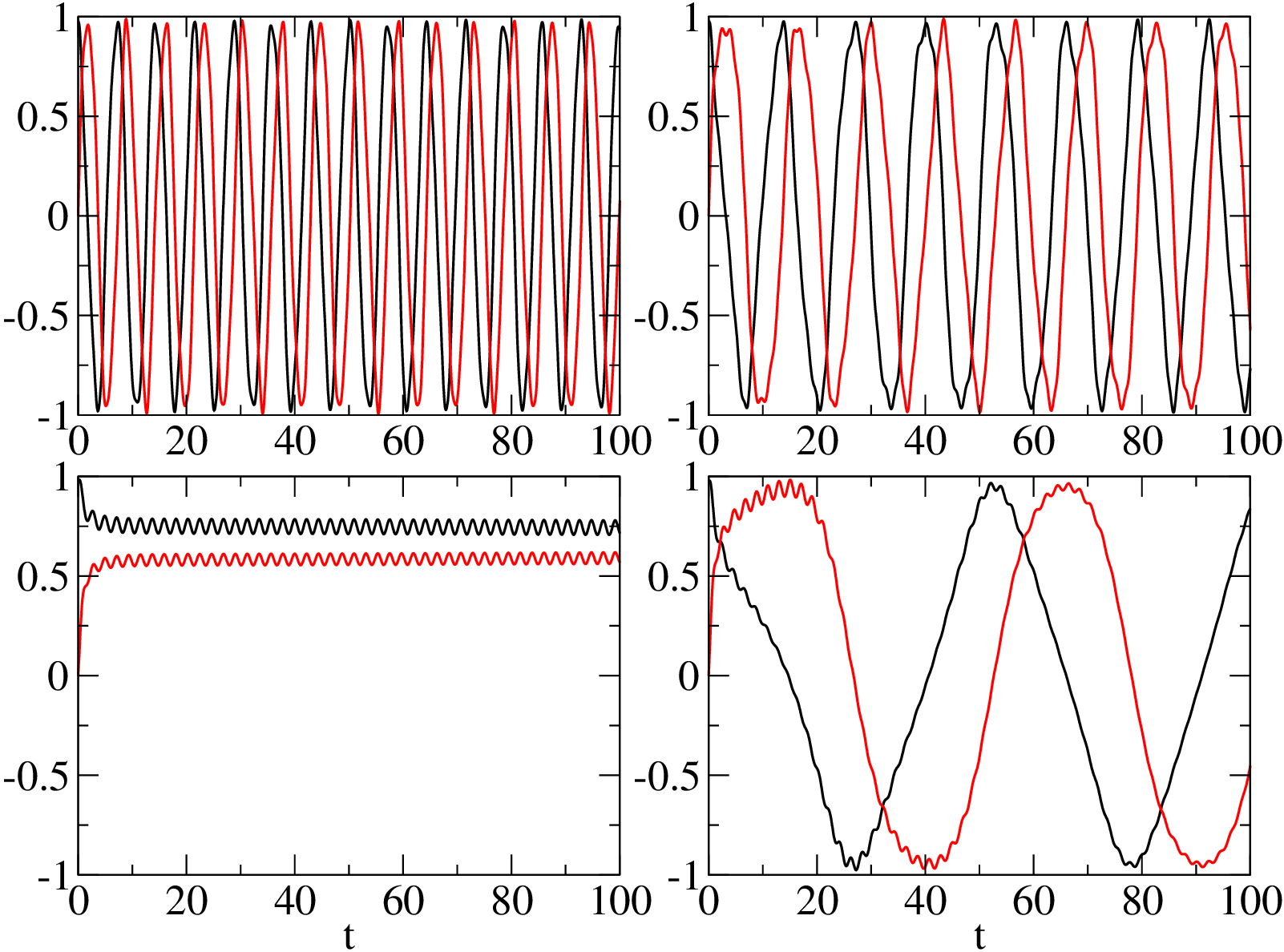}
\caption{(color online) Time evolution of $\Re{(R_{A \uparrow})}$ (black) and $\Im{(R_{A \uparrow})}$ (red) for quenches $U_i=4.0 \rightarrow U_f = 3.2,2.6,2.0,1.6$ (clockwise order from top left). }
\label{fig:R}
\end{figure}

We start by analyzing the dynamics for quenches at $U_f < U_i$. We plot in Fig. \ref{fig:M} the time evolution of the AFM order parameter $m = n_\uparrow - n_\downarrow$ for an interaction quench starting from the optimized variational ground state at $U_i = 4.0$. We immediately recognize a pattern which is very similar to that obtained within DMFT and Hartree-Fock dynamics. \cite{Werner_2} The order parameter $m(t)$ quickly decreases in time after the quench and starts oscillating; as $U_f$ decreases below the critical value of $U_c^{U_f<U_i} \approx 1.7$, the order parameter vanishes. \\
On the same figure we also plot the thermal values $m_{th}$ calculated from the finite temperature Gutzwiller approach\cite{Sandri_finT} at an effective temperature $T_{*}$ such that the equilibrium internal energy is equal to the average energy on the variational wavefunction, which is conserved by the unitary evolution.\\
We note that $m(t)$ oscillates around a value which is more and more distant from the thermal one and stays finite even when $T_{*}$ exceeds the N\'eel temperature, suggesting that the dynamics stays trapped in a non-thermal ordered state in accordance with DMFT result.\cite{Werner_2}
From Fig. \ref{fig:M} two well separated frequencies are distinguishable in the dynamics, which we extract by a discrete Fourier transform and plot in Fig. \ref{fig:spettro}. A high frequency $\omega_1$ sets the fast oscillation and decreases with $U_f$, although staying finite. A lower frequency $\omega_2$ can instead be associated to the presence of magnetic order and vanishes at the critical point as $\propto |U_f-U_c^{U_f<U_i}|$; the existence of a linearly vanishing mode was found also in Ref. \onlinecite{Werner_2}. \\
This two-frequency dynamics reveals the mechanism beyond the disappearance of the AFM order at $U_c^{U_f<U_i}$ . This is more clearly shown in Fig. \ref{fig:R} where we plot the values of the real and imaginary part of the renormalization factors. We observe that approaching $U_c^{U_f<U_i}$ the renormalization factors show main oscillations with frequency $\omega_2$, on top of which there are much  narrower oscillations controlled by $\omega_1$. In proximity of $U_c^{U_f<U_i}$, $\omega_1\gg \omega_2\to 0$, so that, within each ($\bk,\bk+\bQ$) subspace, the magnetic field in the pseudo-spin Hamiltonian (\ref{Hspin}) can be effectively taken constant in time. Hence the dynamics of (\ref{Hspin}) is equivalent to that of a spin in the presence of a $\bk$-dependent constant magnetic field. The total staggered magnetization then vanishes due to the de-phasing that occurs summing on the entire Brillouin zone, hence the nature of the critical point is essentially that found within the Hartree-Fock approximation by Ref. \onlinecite{Werner_2}.\\
Finally, from Fig. \ref{fig:aver} we see that the long time average of $|R_\sigma|^2$ increases in the limit of $U_f\rightarrow 0$, indicating that the AFM insulator actually melts into a PM metal.

\subsection{$U_f > U_i$ quench}
\label{$U_f > U_i$ quench}

\begin{figure}[!h]
\includegraphics[scale=0.32]{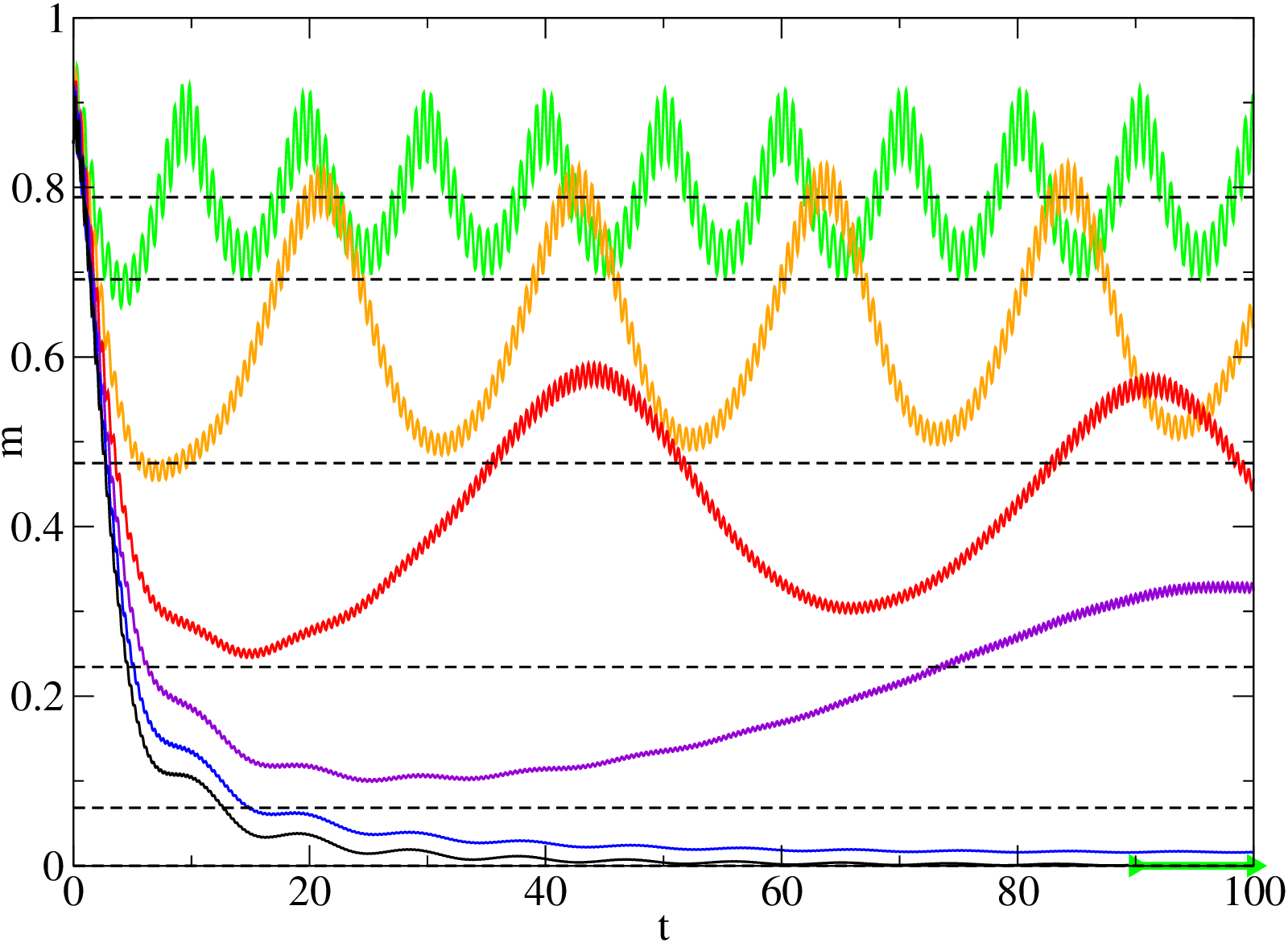}
\caption{(color online) Time evolution of the staggered magnetization $m$ for quenches $U_i=4.0 \rightarrow U_f = 12.0,14.0,16.0,18.0,20.0,22.0$. The green arrow indicates 
the thermal values $m_{th}$ for $U_f=12.0$ and shows that the effective temperature has already crossed the N\'eel temperature. The black dashed lines indicate the values of the long time average. }
\label{fig:mUp}
\end{figure}

\begin{figure}[!h]
\includegraphics[scale=0.32]{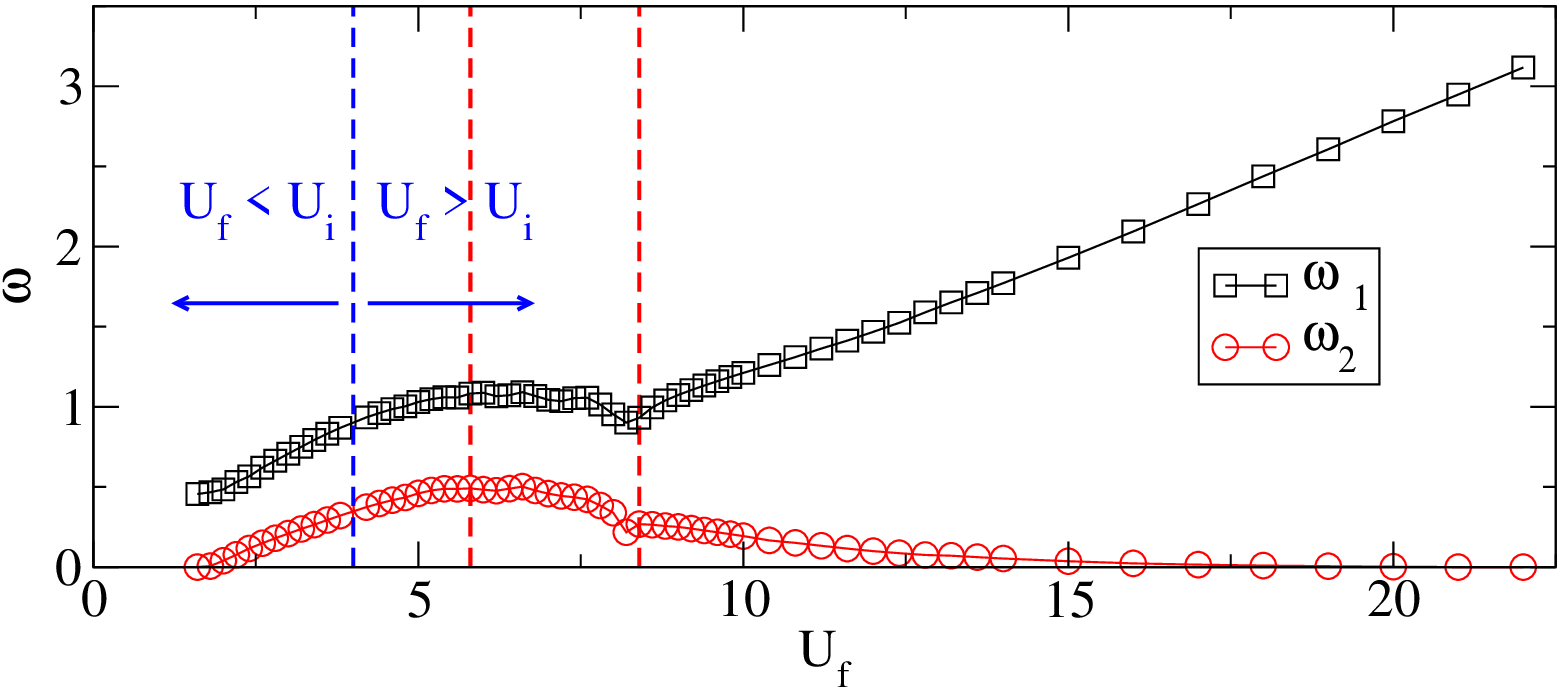}
\caption{(color online) Behaviour of the main drive frequencies $\omega_1$ and $\omega_2$ as a function of $U_f$. The two dashed red lines indicate the crossover region in which the Fourier power spectrum presents broad peaks. }
\label{fig:spettro}
\end{figure}

For quenches at $U_f < U_i$ the Gutzwiller dynamics is not different from the one obtained through single-particle methods such as the Hartree-Fock approximation; the magnetization shows an oscillatory behavior that turns eventually into a fast decay due to dephasing. Differences instead arise when $U_f > U_i$. Here time-dependent Hartree-Fock predicts incorrectly that the magnetic order parameter never vanishes, whatever $U_f$ is. This drawback is directly related to the inadequacy of Hartree-Fock in reproducing a decaying N\'eel temperature at large values of $U$, feature that is instead captured by the Gutzwiller approach, see 
Fig. \ref{fig:phaseFT}. In the assumption that the unitary evolution following the quantum quench brings the 
system in some thermal configuration at finite temperature, the higher the greater $\left|U_f-U_i\right|$, we can not only rationalize why time-dependent Hartree-Fock fails, but also anticipate, within the time-dependent Gutzwiller 
tecnique, a dynamical transition from an antiferromagnetic to a paramagnetic phase. 
Indeed, in the limit of very large $U_f>U_i$,  when the frequency $\omega_1 \sim U_f$ gets much higher than the excitation energies of the Slater determinant, each ($\bk,\bk+\bQ$) pseudo-spin evolves under an effectively  slow magnetic field, hence the staggered magnetization averages again to zero due to dephasing.\\
We find confirmation of this expectation in the time evolution of $m(t)$, see Fig. \ref{fig:mUp}, and the main drive frequencies shown Fig. \ref{fig:spettro}. In the limit of large $U_f$, a two frequency oscillation pattern appears again, with a high frequency $\omega_1$ that grows as $\propto U_f$ and a lower frequency associated with a vanishing mode which decays as $\propto |U_f-U_c^{U_f>U_i}|$ with the critical value of $U_c^{U_f>U_i} \approx 21.0$. \\
We note that also in this regime the long time average of the magnetization differs from the corresponding thermal value. Indeed in Fig. \ref{fig:mUp} we see that for $U_f = 12.0$ the effective temperature has already crossed the N\'eel temperature, while the long time average of the magnetization stays greater than zero, indicating the persistence of a non-equilibrium ordered state in accordance with the results of Ref. \onlinecite{Werner_1}. \\

\begin{figure}[!h]
\includegraphics[scale=0.32]{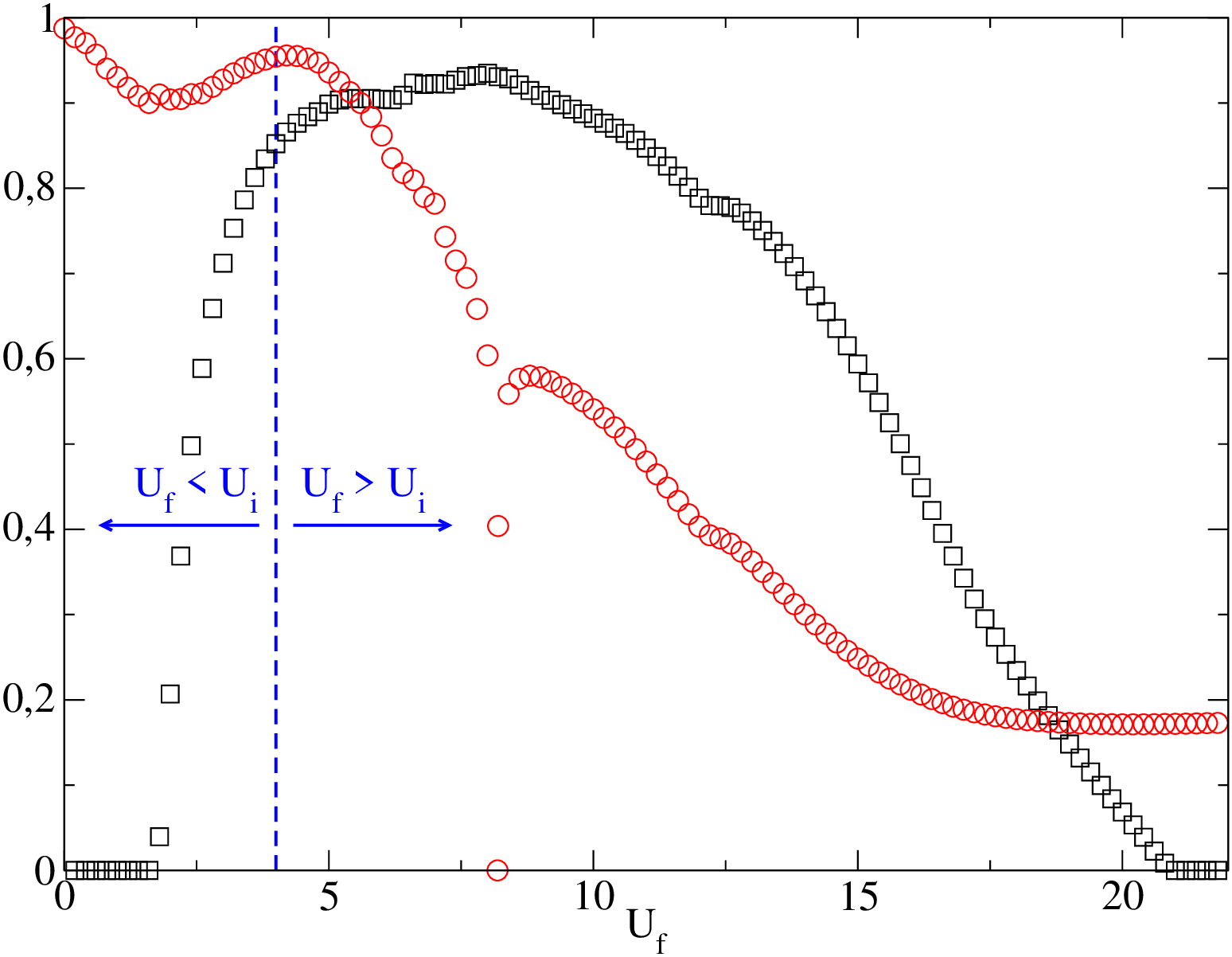}
\caption{(color online) Long time averages of the magnetization (black squares) and of $|R_\sigma|^2$ (red circles) as a function of $U_f$. At $U^c_f\approx8.2$ the renormalization factor time average decays to zero signaling the presence of the dynamical critical point.}
\label{fig:aver}
\end{figure}

\begin{figure}[!h]
\includegraphics[scale=0.32]{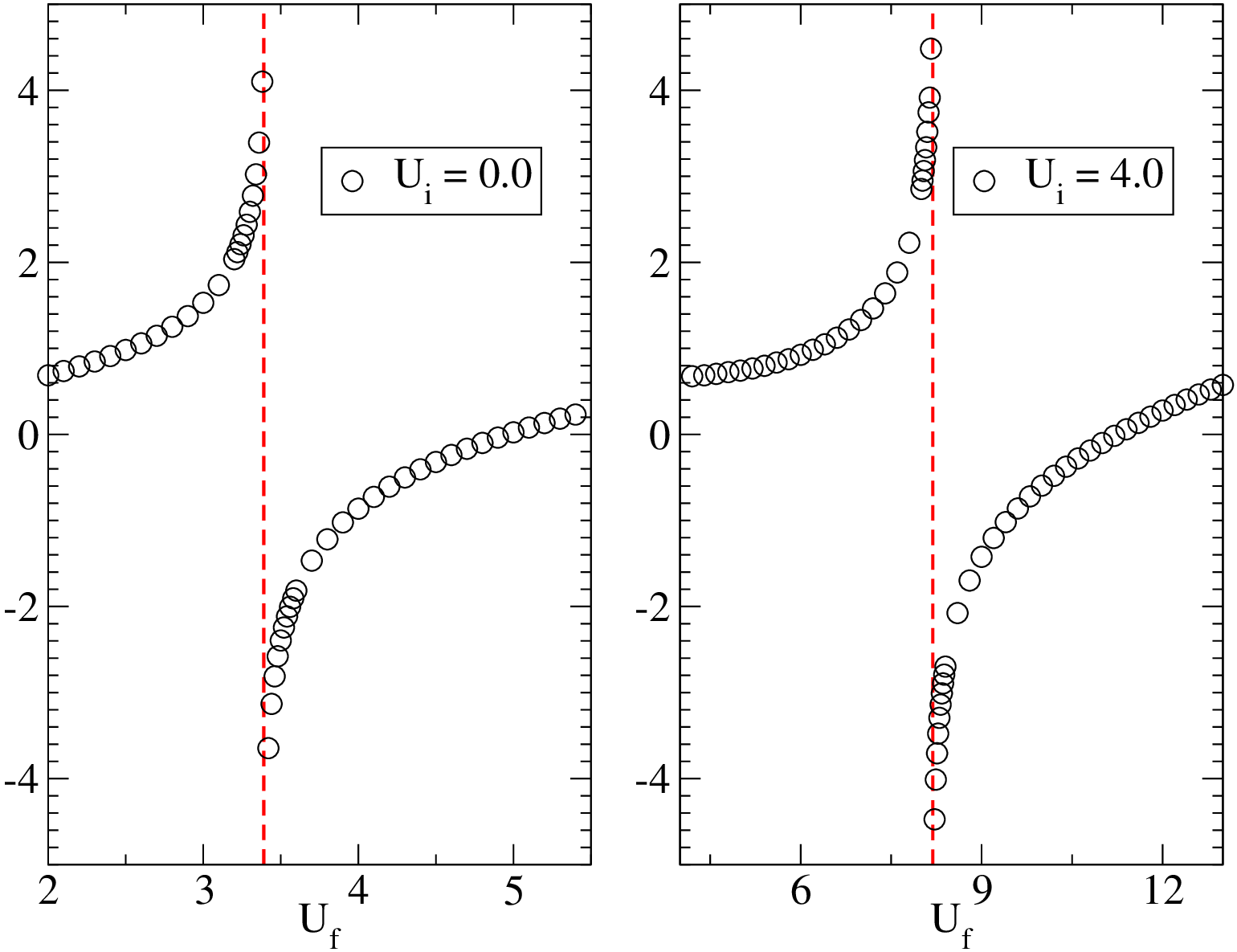}
\caption{(color online) Long time average of $\mathcal{O} = \Re \Big( \frac{\Phi_\uparrow \Phi_\downarrow}{\Phi_0^2} \Big)$ in logarithmic units, i.e. sgn$\bar{\mathcal{O}} \log(|\bar{\mathcal{O}}|) $, for different values of $U_f$. Both in the PM case (left panel) and in the AFM one (right panel) the dynamical critical point is evidenced by a sharp singularity. }
\label{fig:uc}
\end{figure}

For smaller values of $U_f$ instead a less clear scenario appears. Indeed, in the range of values $5.8 \lesssim U_f \lesssim 8.4$ (vertical dashed lines of Fig. \ref{fig:spettro}), although the main frequencies $\omega_1$ and $\omega_2$ can be still recognized by continuity from the large and small $U_f$ limits, the Fourier power spectrum loses regularity and shows an increased number of broad peaks.\\
In this interval of $U_f$,  the long time average of the magnetization increases while 
the renormalization factors diminish, see (Fig. \ref{fig:aver}), suggestive of the systems driven towards a Mott localized regime.\\
We note that Eqs. \eqn{SD1} and \eqn{GP1} admit a stationary solution identified by $R_\sigma=0$ and 
energy equal to zero, which describes a trivial Mott 
insulating state. We find that when the conserved energy after the quench is vanishing, which happens at 
$U_c^{dyn}\approx8.2$ when $U_i=4.0$, Eqs. \eqn{SD1} and \eqn{GP1} flow towards the above stationary solution, see Fig. \ref{fig:aver}, a lot alike what found in the absence of magnetism 
in Ref. \onlinecite{Schiro&Fabrizio_prl}. 
We can shed some light on this dynamical behavior by writing the Gutzwiller parameters as 
\bea
\Phi_{0} &=& \Phi_{\uparrow\downarrow} = \rho_0\, \text{e}^{i \varphi_0},\label{Phi_0}\\
\Phi_\sigma &=& \rho_\sigma \, \text{e}^{i \varphi_\sigma},\label{Phi_sigma}
\eea
with $\rho_{0(\sigma)}\geq 0$ that, because of normalization,  satisfy $2\rho_0^2+\rho_\uparrow^2 
+ \rho_\down^2 =1$ and analyzing the quantity 
\bea 
\Re \bigg( \frac{\Phi_\uparrow \Phi_\downarrow}{\Phi_0^2} \bigg) &=& \fract{\rho_\uparrow \rho_\downarrow}{\rho_0^2}\; \cos\Big(2\varphi_0 - \varphi_\uparrow - \varphi_\downarrow\Big) \nonumber \\
&\equiv&  
\fract{\rho_\uparrow \rho_\downarrow}{\rho_0^2}\; \cos\varphi.
\label{eq:order}
\eea
Neglecting magnetism, which is the same as starting from $U_i=0$, it was shown in Ref. \onlinecite{Schiro&Fabrizio_prl} that the Mott-localized phase can be identified by the dynamics 
of the angle $\varphi$, which reproduces that of a classical pendulum. Below $U_c^{dyn}$, 
$\varphi$ undergoes small oscillations around zero, hence Eq. (\ref{eq:order}) is positive. On the contrary,  above $U_c^{dyn}$,  $\cos \varphi$ starts precessing around the whole unit circle, and, in particular,  is negative right in the regions where the double-occupancy probability $\left|\Phi_{\up\down}\right|^2 = \rho_0^2$ is lower. It follows that, 
for $U_f>U_c^{dyn}$, the quantity in Eq. (\ref{eq:order}) is on average negative. 
Exactly at $U_c^{dyn}$, $\rho_0$ vanishes exponentially, so that the long time average of $\Re \Big( \frac{\Phi_\uparrow \Phi_\downarrow}{\Phi_0^2} \Big)$ diverges and changes sign right at $U_c^{dyn}$, 
see Fig. \ref{fig:uc} left panel. In the right panel of the same figure we show that the same singular behavior  persists also when the system is quenched from an AFM state.  Even though in this case the angle $\varphi$ is not bounded between $[0:2\pi]$ below $U_c^{dyn}$, due to the dynamics of the AFM order parameter, 
yet the time average has a well defined sign that changes crossing a singularity at $U_c^{dyn}$. 

This is suggestive of a dynamical Mott localization at $U_c^{dyn}\approx8.2$, that has  no equilibrium counterpart and separates two different antiferromagnetic insulators. We cannot exclude that this transition may be an artifact of the Gutzwiller technique, although we are tempted to give it a physical meaning. 

In order to clarify this point, we first introduce a more general definition of the 
quasiparticle residue $Z_{\bk \sigma}$ through 
\bea
Z_{\bk \sigma} = |\langle \bk\sigma,N+1| \, c^\dagger_{\bk \sigma} \, |0,N \rangle |^2, \label{def:Z}
\eea 
where $|0,N\rangle$ is the ground state with $N$ electrons, assumed to have zero momentum and spin,  and $|\bk\sigma,N+1\rangle$ the lowest energy 
state with $N+1$ electrons, momentum $\bk$ and spin $\sigma$. $Z_{\bk\sigma}$ defined by 
Eq. \eqn{def:Z} coincides with the jump of the momentum distribution at the Fermi surface $|\bk|=k_F$ 
for a Landau-Fermi liquid, but remains well defined also for an insulator, where it can be used to establish whether well-defined quasiparticles exist above the gap.  Indeed, one can readily realize that 
$Z_{\bk\sigma}=1$ for a non-interacting band-insulator. Therefore, one can in principle distinguish two 
different insulators: a ``coherent'' insulator akin to a band insulator with $0<Z_{\bk\sigma}\leq 1$, 
and an ``incoherent'' insulator, similar to an idealized Mott insulator, with $Z_{\bk\sigma}=0$ and no well-defined quasiparticles above the gap. 

We then observe that, at zero temperature, $|R_{\sigma}|^2$ defined by Eq. \eqn{RENO} is just an estimate, within the Gutzwiller approximation, of $Z_{\bk \sigma}$ above.    
Indeed, one can readily prove that 
\bea
\langle \bk\sigma,N+1| \, c^\dagger_{\bk \sigma}\, |0,N \rangle  &\overset{GW}{=}& \langle \psi_N|\,  c_{\bk \sigma} \; \m{P}\, c^\dagger_{\bk \sigma} \,\m{P}\, | \psi_N \rangle \nonumber\\
&=& R_\sigma .\label{Z-GA}
\eea
Here we used the fact that the Gutzwiller wavefunction $\mathcal{P}\, |\psi_N \rangle$ (with $|\psi_N \rangle$ the $N$-particle Slater determinant that defines the variational wave function in Eq. \eqn{GWF}) is the variational estimate of $|0,N \rangle$ and that, within corrections 
$O(N^{-1})$, the best variational estimate of the $(N+1)$-electron lowest energy wave function with momentum $\bk$ and spin $\sigma$ is just $|\bk\sigma,N+1\rangle \simeq \m{P}\,c^\dagger_{\bk\sigma}\,|\Psi_N\rangle$, with the same $\m{P}$ as 
for $N$ electrons. Eq. \eqn{Z-GA} remains valid also in the time dependent case where the evolution of the ground state, being a pure state, is approximated by Eq. (\ref{GWF}).

We thence arrive to the conclusion that our dynamical transition separates two different antiferromagnetic insulators in the above meaning, one characterized by a finite $Z$ and the other by a vanishing one. It is worth mentioning that at equilibrium and zero temperature, all evidences indicate that $Z$ of Eq. \eqn{def:Z} is everywhere finite in the antiferromagnetic insulating phase of the Hubbard model at any value of $U$, as confirmed by DMFT\cite{Kotliar-Chitra} and by quantum Monte Carlo simulations on the $t$-$J$ model.\cite{Muramatsu} In other words, even at very large $U$ where the Mott's physics dominates and local moments are already well formed, the antiferromagnet has coherent quasiparticles above the gap. We actually believe that, as soon as 
long-range magnetic order sets in below the N\'eel temperature, the quasiparticle residue $Z$ becomes finite 
at equilibrium. In fact, the onset of long-range order is accompanied at large $U$ by a hopping energy gain, through the spin-exchange $t^2/U$, hence by a raise of lattice coherence that we think has to be associated with an increase of $Z$. That is why we think that the dynamical transition that we observe has no equilibrium counterpart in the whole $U$ versus temperature phase diagram.

We conclude mentioning that the main results presented above at fixed $U_i=4$, remain 
qualitatively the same also at different $U_i$. We indeed verified the presence of the critical points at which the magnetization vanishes, $U_c^{U_f \lessgtr U_i}$, and the presence of the dynamical critical point, $U_i<U_c^{dyn}<U_c^{U_f > U_i}$, for all values of $U_i < 10.0$.  

\section{Concluding remarks}
\label{Concluding remarks}

We have shown that the time dependent Gutzwiller technique, in spite of its simplicity, is able to reproduce the main features of a quench dynamics from an antiferromagnetic state found by time-dependent DMFT, such as the existence of non-thermal magnetically ordered states that disappears above dynamical critical points, both suddenly decreasing or increasing the value of the Hubbard $U$. In addition, we have found evidence of  an additional dynamical transition that occurs at large $U$,  which we interpret as a dynamical Mott transition separating two different antiferromagnetic non-equilibrium states, one characterized by a finite quasiparticle residue and the other by a vanishing one. Since the quasiparticle residue $Z$ in an antiferromagnet cannot be extracted by any static property (unlike in a paramagnet where, at zero temperature, $Z$ is the jump of the momentum distribution at the Fermi surface), but requires calculating for instance the full out-of-equilibrium self-energy, its dynamical behavior was not addressed by DMFT in Ref. 
\onlinecite{Werner_1} and Ref. \onlinecite{Werner_2}.  Although we cannot exclude that the vanishing of $Z$ that we observe could be an artifact of the Gutzwiller technique, nevertheless this result is intriguing, as it entails the 
existence in out-of-equilibrium of an incoherent antiferromagnet,  hence worth to be further investigated. 

\acknowledgments
We thank Adriano Amaricci for very helpful suggestions.
This work has been supported by 
the European Union, Seventh Framework Programme, under the project
GO FAST, grant agreement no. 280555.


\begin{thebibliography}{19}
\expandafter\ifx\csname natexlab\endcsname\relax\def\natexlab#1{#1}\fi
\expandafter\ifx\csname bibnamefont\endcsname\relax
  \def\bibnamefont#1{#1}\fi
\expandafter\ifx\csname bibfnamefont\endcsname\relax
  \def\bibfnamefont#1{#1}\fi
\expandafter\ifx\csname citenamefont\endcsname\relax
  \def\citenamefont#1{#1}\fi
\expandafter\ifx\csname url\endcsname\relax
  \def\url#1{\texttt{#1}}\fi
\expandafter\ifx\csname urlprefix\endcsname\relax\def\urlprefix{URL }\fi
\providecommand{\bibinfo}[2]{#2}
\providecommand{\eprint}[2][]{\url{#2}}

\bibitem[{\citenamefont{Bloch et~al.}(2008)\citenamefont{Bloch, Dalibard, and
  Zwerger}}]{Bloch}
\bibinfo{author}{\bibfnamefont{I.}~\bibnamefont{Bloch}},
  \bibinfo{author}{\bibfnamefont{J.}~\bibnamefont{Dalibard}}, \bibnamefont{and}
  \bibinfo{author}{\bibfnamefont{W.}~\bibnamefont{Zwerger}},
  \bibinfo{journal}{Rev. Mod. Phys.} \textbf{\bibinfo{volume}{80}},
  \bibinfo{pages}{885} (\bibinfo{year}{2008}),
  \urlprefix\url{http://link.aps.org/doi/10.1103/RevModPhys.80.885}.

\bibitem[{\citenamefont{Cavalieri et~al.}(2007)\citenamefont{Cavalieri, Muller,
  Uphues, Yakovlev, Baltuska, Horvath, Blumel, Holzwarth, Hendel, Drescher
  et~al.}}]{Cavalieri}
\bibinfo{author}{\bibfnamefont{A.~L.} \bibnamefont{Cavalieri}},
  \bibinfo{author}{\bibfnamefont{N.}~\bibnamefont{Muller}},
  \bibinfo{author}{\bibfnamefont{T.}~\bibnamefont{Uphues}},
  \bibinfo{author}{\bibfnamefont{V.~S.} \bibnamefont{Yakovlev}},
  \bibinfo{author}{\bibfnamefont{A.}~\bibnamefont{Baltuska}},
  \bibinfo{author}{\bibfnamefont{B.}~\bibnamefont{Horvath},
  \bibfnamefont{Schmidt}},
  \bibinfo{author}{\bibfnamefont{L.}~\bibnamefont{Blumel}},
  \bibinfo{author}{\bibfnamefont{R.}~\bibnamefont{Holzwarth}},
  \bibinfo{author}{\bibfnamefont{S.}~\bibnamefont{Hendel}},
  \bibinfo{author}{\bibfnamefont{M.}~\bibnamefont{Drescher}},
  \bibnamefont{et~al.}, \bibinfo{journal}{Nature}
  \textbf{\bibinfo{volume}{449}}, \bibinfo{pages}{1029} (\bibinfo{year}{2007}).

\bibitem[{\citenamefont{Perfetti et~al.}(2006)\citenamefont{Perfetti, Loukakos,
  Lisowski, Bovensiepen, Berger, Biermann, Cornaglia, Georges, and
  Wolf}}]{Perfetti}
\bibinfo{author}{\bibfnamefont{L.}~\bibnamefont{Perfetti}},
  \bibinfo{author}{\bibfnamefont{P.~A.} \bibnamefont{Loukakos}},
  \bibinfo{author}{\bibfnamefont{M.}~\bibnamefont{Lisowski}},
  \bibinfo{author}{\bibfnamefont{U.}~\bibnamefont{Bovensiepen}},
  \bibinfo{author}{\bibfnamefont{H.}~\bibnamefont{Berger}},
  \bibinfo{author}{\bibfnamefont{S.}~\bibnamefont{Biermann}},
  \bibinfo{author}{\bibfnamefont{P.~S.} \bibnamefont{Cornaglia}},
  \bibinfo{author}{\bibfnamefont{A.}~\bibnamefont{Georges}}, \bibnamefont{and}
  \bibinfo{author}{\bibfnamefont{M.}~\bibnamefont{Wolf}},
  \bibinfo{journal}{Phys. Rev. Lett.} \textbf{\bibinfo{volume}{97}},
  \bibinfo{pages}{067402} (\bibinfo{year}{2006}),
  \urlprefix\url{http://link.aps.org/doi/10.1103/PhysRevLett.97.067402}.

\bibitem[{\citenamefont{Beaud et~al.}(2009)\citenamefont{Beaud, Johnson,
  Vorobeva, Staub, Souza, Milne, Jia, and Ingold}}]{Beaud}
\bibinfo{author}{\bibfnamefont{P.}~\bibnamefont{Beaud}},
  \bibinfo{author}{\bibfnamefont{S.~L.} \bibnamefont{Johnson}},
  \bibinfo{author}{\bibfnamefont{E.}~\bibnamefont{Vorobeva}},
  \bibinfo{author}{\bibfnamefont{U.}~\bibnamefont{Staub}},
  \bibinfo{author}{\bibfnamefont{R.~A.~D.} \bibnamefont{Souza}},
  \bibinfo{author}{\bibfnamefont{C.~J.} \bibnamefont{Milne}},
  \bibinfo{author}{\bibfnamefont{Q.~X.} \bibnamefont{Jia}}, \bibnamefont{and}
  \bibinfo{author}{\bibfnamefont{G.}~\bibnamefont{Ingold}},
  \bibinfo{journal}{Phys. Rev. Lett.} \textbf{\bibinfo{volume}{103}},
  \bibinfo{pages}{155702} (\bibinfo{year}{2009}),
  \urlprefix\url{http://link.aps.org/doi/10.1103/PhysRevLett.103.155702}.

\bibitem[{\citenamefont{Fausti et~al.}(2011)\citenamefont{Fausti, Tobey, Dean,
  Kaiser, Dienst, Hoffmann, Pyon, Takayama, Takagi, and Cavalleri}}]{Fausti}
\bibinfo{author}{\bibfnamefont{D.}~\bibnamefont{Fausti}},
  \bibinfo{author}{\bibfnamefont{R.~I.} \bibnamefont{Tobey}},
  \bibinfo{author}{\bibfnamefont{N.}~\bibnamefont{Dean}},
  \bibinfo{author}{\bibfnamefont{S.}~\bibnamefont{Kaiser}},
  \bibinfo{author}{\bibfnamefont{A.}~\bibnamefont{Dienst}},
  \bibinfo{author}{\bibfnamefont{M.~C.} \bibnamefont{Hoffmann}},
  \bibinfo{author}{\bibfnamefont{S.}~\bibnamefont{Pyon}},
  \bibinfo{author}{\bibfnamefont{T.}~\bibnamefont{Takayama}},
  \bibinfo{author}{\bibfnamefont{H.}~\bibnamefont{Takagi}}, \bibnamefont{and}
  \bibinfo{author}{\bibfnamefont{A.}~\bibnamefont{Cavalleri}},
  \bibinfo{journal}{Science} \textbf{\bibinfo{volume}{331}},
  \bibinfo{pages}{189} (\bibinfo{year}{2011}),
  \eprint{http://www.sciencemag.org/content/331/6014/189.full.pdf},
  \urlprefix\url{http://www.sciencemag.org/content/331/6014/189.abstract}.

\bibitem[{\citenamefont{Mazza and Fabrizio}(2012)}]{Giacomo}
\bibinfo{author}{\bibfnamefont{G.}~\bibnamefont{Mazza}} \bibnamefont{and}
  \bibinfo{author}{\bibfnamefont{M.}~\bibnamefont{Fabrizio}},
  \bibinfo{journal}{Phys. Rev. B} \textbf{\bibinfo{volume}{86}},
  \bibinfo{pages}{184303} (\bibinfo{year}{2012}),
  \urlprefix\url{http://link.aps.org/doi/10.1103/PhysRevB.86.184303}.

\bibitem[{\citenamefont{Werner et~al.}(2012)\citenamefont{Werner, Tsuji, and
  Eckstein}}]{Werner_1}
\bibinfo{author}{\bibfnamefont{P.}~\bibnamefont{Werner}},
  \bibinfo{author}{\bibfnamefont{N.}~\bibnamefont{Tsuji}}, \bibnamefont{and}
  \bibinfo{author}{\bibfnamefont{M.}~\bibnamefont{Eckstein}},
  \bibinfo{journal}{Phys. Rev. B} \textbf{\bibinfo{volume}{86}},
  \bibinfo{pages}{205101} (\bibinfo{year}{2012}),
  \urlprefix\url{http://link.aps.org/doi/10.1103/PhysRevB.86.205101}.

\bibitem[{\citenamefont{Tsuji et~al.}(2013)\citenamefont{Tsuji, Eckstein, and
  Werner}}]{Werner_2}
\bibinfo{author}{\bibfnamefont{N.}~\bibnamefont{Tsuji}},
  \bibinfo{author}{\bibfnamefont{M.}~\bibnamefont{Eckstein}}, \bibnamefont{and}
  \bibinfo{author}{\bibfnamefont{P.}~\bibnamefont{Werner}},
  \bibinfo{journal}{Phys. Rev. Lett.} \textbf{\bibinfo{volume}{110}},
  \bibinfo{pages}{136404} (\bibinfo{year}{2013}),
  \urlprefix\url{http://link.aps.org/doi/10.1103/PhysRevLett.110.136404}.

\bibitem[{\citenamefont{Schir\'o and Fabrizio}(2010)}]{Schiro&Fabrizio_prl}
\bibinfo{author}{\bibfnamefont{M.}~\bibnamefont{Schir\'o}} \bibnamefont{and}
  \bibinfo{author}{\bibfnamefont{M.}~\bibnamefont{Fabrizio}},
  \bibinfo{journal}{Phys. Rev. Lett.} \textbf{\bibinfo{volume}{105}},
  \bibinfo{pages}{076401} (\bibinfo{year}{2010}),
  \urlprefix\url{http://link.aps.org/doi/10.1103/PhysRevLett.105.076401}.

\bibitem[{\citenamefont{Schir\'o and Fabrizio}(2011)}]{Schiro&Fabrizio_prb}
\bibinfo{author}{\bibfnamefont{M.}~\bibnamefont{Schir\'o}} \bibnamefont{and}
  \bibinfo{author}{\bibfnamefont{M.}~\bibnamefont{Fabrizio}},
  \bibinfo{journal}{Phys. Rev. B} \textbf{\bibinfo{volume}{83}},
  \bibinfo{pages}{165105} (\bibinfo{year}{2011}),
  \urlprefix\url{http://link.aps.org/doi/10.1103/PhysRevB.83.165105}.

\bibitem[{\citenamefont{Sandri et~al.}(2012)\citenamefont{Sandri, Schir\'o, and
  Fabrizio}}]{Sandri_lin}
\bibinfo{author}{\bibfnamefont{M.}~\bibnamefont{Sandri}},
  \bibinfo{author}{\bibfnamefont{M.}~\bibnamefont{Schir\'o}}, \bibnamefont{and}
  \bibinfo{author}{\bibfnamefont{M.}~\bibnamefont{Fabrizio}},
  \bibinfo{journal}{Phys. Rev. B} \textbf{\bibinfo{volume}{86}},
  \bibinfo{pages}{075122} (\bibinfo{year}{2012}),
  \urlprefix\url{http://link.aps.org/doi/10.1103/PhysRevB.86.075122}.

\bibitem[{\citenamefont{Fabrizio}(2013)}]{Hvar}
\bibinfo{author}{\bibfnamefont{M.}~\bibnamefont{Fabrizio}}, in
  \emph{\bibinfo{booktitle}{New Materials for Thermoelectric Applications:
  Theory and Experiment}}, edited by
  \bibinfo{editor}{\bibfnamefont{V.}~\bibnamefont{Zlatic}} \bibnamefont{and}
  \bibinfo{editor}{\bibfnamefont{A.}~\bibnamefont{Hewson}}
  (\bibinfo{publisher}{Springer}, \bibinfo{year}{2013}), NATO Science for Peace
  and Security Series - B: Physics and Biophysics, pp.
  \bibinfo{pages}{247--272}.

\bibitem[{\citenamefont{Sandri et~al.}(2013)\citenamefont{Sandri, Capone, and
  Fabrizio}}]{Sandri_finT}
\bibinfo{author}{\bibfnamefont{M.}~\bibnamefont{Sandri}},
  \bibinfo{author}{\bibfnamefont{M.}~\bibnamefont{Capone}}, \bibnamefont{and}
  \bibinfo{author}{\bibfnamefont{M.}~\bibnamefont{Fabrizio}},
  \bibinfo{journal}{Phys. Rev. B} \textbf{\bibinfo{volume}{87}},
  \bibinfo{pages}{205108} (\bibinfo{year}{2013}),
  \urlprefix\url{http://link.aps.org/doi/10.1103/PhysRevB.87.205108}.

\bibitem[{\citenamefont{Hamerla and Uhrig}(2013{\natexlab{a}})}]{Hamerla1}
\bibinfo{author}{\bibfnamefont{S.~A.} \bibnamefont{Hamerla}} \bibnamefont{and}
  \bibinfo{author}{\bibfnamefont{G.~S.} \bibnamefont{Uhrig}},
  \bibinfo{journal}{Phys. Rev. B} \textbf{\bibinfo{volume}{87}},
  \bibinfo{pages}{064304} (\bibinfo{year}{2013}{\natexlab{a}}),
  \urlprefix\url{http://link.aps.org/doi/10.1103/PhysRevB.87.064304}.

\bibitem[{\citenamefont{Hamerla and Uhrig}(2013{\natexlab{b}})}]{Hamerla2}
\bibinfo{author}{\bibfnamefont{S.~A.} \bibnamefont{Hamerla}} \bibnamefont{and}
  \bibinfo{author}{\bibfnamefont{G.~S.} \bibnamefont{Uhrig}}
  (\bibinfo{year}{2013}{\natexlab{b}}), \bibinfo{note}{arXiv:1307.3438}.

\bibitem[{\citenamefont{Lanat\`a et~al.}(2008)\citenamefont{Lanat\`a, Barone,
  and Fabrizio}}]{Lanata-Kondo}
\bibinfo{author}{\bibfnamefont{N.}~\bibnamefont{Lanat\`a}},
  \bibinfo{author}{\bibfnamefont{P.}~\bibnamefont{Barone}}, \bibnamefont{and}
  \bibinfo{author}{\bibfnamefont{M.}~\bibnamefont{Fabrizio}},
  \bibinfo{journal}{Phys. Rev. B} \textbf{\bibinfo{volume}{78}},
  \bibinfo{pages}{155127} (\bibinfo{year}{2008}),
  \urlprefix\url{http://link.aps.org/doi/10.1103/PhysRevB.78.155127}.

\bibitem[{\citenamefont{Lanat\`a et~al.}(2013)\citenamefont{Lanat\`a, Yao,
  Wang, Ho, Schmalian, Haule, and Kotliar}}]{Lanata-Cerio}
\bibinfo{author}{\bibfnamefont{N.}~\bibnamefont{Lanat\`a}},
  \bibinfo{author}{\bibfnamefont{Y.-X.} \bibnamefont{Yao}},
  \bibinfo{author}{\bibfnamefont{C.-Z.} \bibnamefont{Wang}},
  \bibinfo{author}{\bibfnamefont{K.-M.} \bibnamefont{Ho}},
  \bibinfo{author}{\bibfnamefont{J.}~\bibnamefont{Schmalian}},
  \bibinfo{author}{\bibfnamefont{K.}~\bibnamefont{Haule}}, \bibnamefont{and}
  \bibinfo{author}{\bibfnamefont{G.}~\bibnamefont{Kotliar}}
  (\bibinfo{year}{2013}), \bibinfo{note}{arXiv:1305.3950}.

\bibitem[{\citenamefont{Chitra and Kotliar}(1999)}]{Kotliar-Chitra}
\bibinfo{author}{\bibfnamefont{R.}~\bibnamefont{Chitra}} \bibnamefont{and}
  \bibinfo{author}{\bibfnamefont{G.}~\bibnamefont{Kotliar}},
  \bibinfo{journal}{Phys. Rev. Lett.} \textbf{\bibinfo{volume}{83}},
  \bibinfo{pages}{2386} (\bibinfo{year}{1999}),
  \urlprefix\url{http://link.aps.org/doi/10.1103/PhysRevLett.83.2386}.

\bibitem[{\citenamefont{Brunner et~al.}(2000)\citenamefont{Brunner, Assaad, and
  Muramatsu}}]{Muramatsu}
\bibinfo{author}{\bibfnamefont{M.}~\bibnamefont{Brunner}},
  \bibinfo{author}{\bibfnamefont{F.~F.} \bibnamefont{Assaad}},
  \bibnamefont{and}
  \bibinfo{author}{\bibfnamefont{A.}~\bibnamefont{Muramatsu}},
  \bibinfo{journal}{Phys. Rev. B} \textbf{\bibinfo{volume}{62}},
  \bibinfo{pages}{15480} (\bibinfo{year}{2000}),
  \urlprefix\url{http://link.aps.org/doi/10.1103/PhysRevB.62.15480}.

\end{thebibliography}
\end{document}